\documentclass[final]{article}
\usepackage[latin9]{inputenc}
\usepackage{array}
\usepackage{amsmath}
\usepackage{amssymb}

\makeatletter

\providecommand{\tabularnewline}{\\}

\newcommand{\lyxaddress}[1]{
\par {\raggedright #1
\vspace{1.4em}
\noindent\par}
}

\newcommand{\boldmathPsi}{\mbox{\boldmath$\Psi$\unboldmath}}

\usepackage{array}
\usepackage{graphicx}
\usepackage{tikz}
\usepackage[figuresright]{rotating}
\usepackage{adjustbox}

\providecommand{\keywords}[1]{\textbf{\textit{Keywords:}} #1}

\date{\vspace{-5ex}}

\@ifundefined{showcaptionsetup}{}{%
 \PassOptionsToPackage{caption=false}{subfig}}
\usepackage{subfig}
\makeatother

\begin{document}

\title{Particles within extended-spin space: Lagrangian connection }

\author{J. Besprosvany and R. Romero}

\maketitle

\lyxaddress{Instituto de F\'{\i}sica, Universidad Nacional Aut\'onoma de M\'exico,
Apartado Postal 20-364, M\'exico 01000, Distrito  Federal, M\'exico }
\begin{abstract}
A spin-space extension is reviewed, which provides information on
the standard model. Its defining feature is a common matrix space
that describes symmetries and representations, and leads to limits
on these, for given dimension. The model provides additional information
on the standard model, whose interpretation requires an interactive
formulation. Within this program, we compare the model's lepton-W
generated interactive Lagrangian in (5+1)-dimensions, and that of
the standard model. We derive the conditions for this matching, which
apply to other Lagrangian terms. We also discuss the advantages of
this extension, as compared to others.
\end{abstract}

\section{Introduction}

\subsection{Background}

The standard model is the theory that describes the key elements of
nature, also one of the most successful theories and, at the same
time, which presents the greatest enigmas in modern physics. Although
the model correctly describes the elementary particles, it is phenomenological.
On the one hand, the fermion representations have been established,
as well as their classification in generations, and the forces acting
between these particles, which define the vector bosons transmitting
these interactions. On the other hand, the origin of the specific
types of representations and forces that nature has chosen is not
known. In particular, we do not know why matter consists of leptons
and quarks, nor the reason for the interaction groups U$_{Y}$(1)$\times$SU$_{L}$(2)$\times$SU(3)
and related particles: the Z boson carries the hypercharge, is associated
to the group U$_{Y}$(1), and applies to all particles; the W bosons
are associated to the SU(2)$_{L}$ group and act on left-chirality
particles; gluons produce the strong interaction, derive from the
color group SU(3), and act only upon the quarks. The origin of this
behavior is unknown. Finally, we need a more fundamental reason for
the existence of the scalar particle that is suggested in recent experiments\cite{CMS,Atlas},
the Higgs, and which gives mass to particles. This ignorance is also
reflected in the relative large number of parameters required by the
model, of the order of twenty, as the particle masses and charges,
which are fixed experimentally. By the nature of the standard model,
it is understood that by itself, it will never explain these unknowns
and that, therefore, we need to investigate options beyond it.

Great insights have been reached throughout the history of Physics
by the discovery of connections between phenomena. Traditional examples
include Newton's connection of the Moon's movement with the fall of
an object on Earth, through gravity, and Maxwell's understanding of
light as an electromagnetic phenomenon, obtained from wave solutions
in his equations, and the speed of light built in terms of the relative
permittivity and magnetic permeability of the vacuum. Furthermore,
advances in the understanding of elementary particles have been obtained
from a framework that assumes unification and/or symmetry of the above
physical quantities that characterize them. Indeed, the successes
of the past include the chiral-symmetry assumption, involved in the
generation of hadron masses; supersymmetry, a hypothesis currently
under investigation, explains symmetry breaking at the low-energy
electroweak scale, and creates the masses of the known elementary
particles.

A partial but practical description of the fundamental physical elements
that participate in the modern unification ideas consists of particles,
classified as bosons and fermions; spin and space as their associated
attributes; and finally, their interactions, as described within general
relativity and the standard model. These are the key elements to investigate.
Before introducing this paper's proposal, we briefly review some standard-model
extensions:

\subsection{Kaluza-Klein and grand-unification theories}

A promising unification is the idea of Kaluza-Klein, who proposed
extra spatial dimensions{[}d{]}, beyond 3 + 1, to be associated with
gauge symmetries. In the case of grand-unification theories in their
application to the standard model, there are restrictions on the standard-model
U(1)$_{Y}$$\times$ SU(2)$_{L}$$\times$SU(3) gauge groups, as well
as on the representations and coupling-constant values.

\subsection{Quasi-particles}

This idea, originated by Landau, suggests that it is possible to achieve
an adequate description of interactive particles, if one manages to
describe their effective degrees of freedom in an appropriate way.
To first order, it would be possible to consider particles as free,
while parameters such as mass would be modified. The search for these
degrees of freedom represents one of the main objectives in studies
in areas that engage many-particle systems, with quantum behavior,
such as nuclear physics and superconductivity. Indeed, in the area
of elementary particles, Nambu and Jona-Lasinio\cite{Nambu} described
an interactive model within the framework of field theory, inspired
by superconductivity, and which leads to masses of composite particles,
from an assumed interaction. The lesson is that finding the correct
degrees of freedom may hold the clue to gain insight into the standard
model.

\subsection{Extended-spin model}

As for the actual description of the elementary particles, we concentrate
on their degrees of freedom. Particles and interactions obey Lorentz
and scalar symmetries, global and local, and are described with non-trivial
discrete quantum numbers. While the space degree of freedom is common
to all elementary particles, the discrete degrees of freedom associated
with the fundamental representations are more elementary insofar as
they can be used to build the others. Spin is a physical manifestation
of the representation of the Lorentz group. In relation to space,
spin maintains this role since the first uses the vector representation,
and can be constructed in terms of the second. Other similar investigations
underscore the spin degree of freedom in the extensions of the standard
model (see, e. g., Refs. \cite{Shima:2000es,Chisholm:1999pg,raey}).
The fact that the known fermions participate in the fundamental representation
of the Lorentz and gauge groups, and that the gauge bosons, the interaction
carriers, belong to the adjoint representation of these groups, suggests
a common description\cite{Besprosvany:2002zr}. Indeed, such similarities
and the presence of symmetry suggests a unified description, i.e.,
an elementary space for the discrete degrees of freedom: Lorentz and
scalar. In fact, there are similar common requirements that emerge
from the quantum description and quantization of particles and interactions,
such as the restrictions on representations from unitarity.

The extended-spin model, just as the idea of Kaluza-Klein, assumes
a common space for the spin and scalar degrees of freedom. While the
idea of mixing these is tempting, the Coleman and Mandula theorem\cite{Coleman:1967ad}
prohibits a non-trivial mixing. Obeying this restriction means that
the resulting scalar generators commute with the Lorentz ones, which
is equivalent to the requirement that these two elements be described
as direct products. However, a simple classification of spaces is
permitted with symmetries as the chiral one, and this leads to limitations
in the elements that can be obtained within the space, which ultimately,
gives information, for example, on representations and interactions.
New information is derived as constraints on the chirality of the
interactions and representationss\cite{Besprosvany:2002zr,Besprosvany2001,Besprosvany:2002tv},
the coupling constants\cite{Besprosvany:2002zr,Besprosvany:2002tv,Besprosvany:2002py},
connections among the standard-model particle masses\cite{Besprosvany:2002py},
and a fermion hierarchy effect\cite{Besprosvany:2014lwa}.

The spin-extended model can be interpreted within the Kaluza-Klein
framework, as a result that the additional spatial dimensional components
are frozen. Conceptually, the construction of the model in terms of
matrices comes from incremental direct products with 2$\times$2 matrices,
suggesting the discrete Hilbert space considered is made from elementary
degrees of freedom (e. g., q-bits or particles of spin 1/2).

A field theory can be equivalently formulated in terms of such degrees
of freedom. Work on that direction was carried out on Ref. \cite{Besprosvany:2014lwa}.
In this paper, after introducing the spin-extended model by presenting
its landmarks, we examine in detail its formulation within a standard
Lagrangian, using representation fields and symmetries that derive
from it; in particular, we look at a specific vertex and study its
connection to a standard formulation. This complements Ref. \cite{Besprosvany:2014lwa},
which also deals with this connection, with a general analysis of
the fields' construction, various vertices, and symmetry implementation.
Here we examine the W-fermion interaction term derived from (5+1)-d,
making a detailed description of its Lagrangian, with further analysis
of the projection operator involved, allowing for this equivalence.
In particular, we focus on its coefficients and phases, extending
previous work \cite{Besprosvany:2002py,Besprosvany:2014lwa,Besprosvany:2010zz}.
The dimension $N=4$ case was analyzed in Refs. \cite{Besprosvany:2002zr,Besprosvany2001},
$N=6$ in \cite{Besprosvany:2002zr}, \cite{Besprosvany2001}, \cite{Besprosvany:2002py},
$N=8$ partially in \cite{Besprosvany:2014lwa}, and $N=10$ in \cite{Besprosvany:2002tv}.

The paper is organized as follows: In Section 2, we review the construction
of the proposed extended-spin space, based on a matrix space. For
this purpose, we present as example a massless Hamiltonian. A Clifford
algebra helps in the classification of both operators and states.
Under the demand that the Lorentz symmetry be maintained, scalar degrees
of freedom emerge, associated to global and gauge symmetries. The
matrix space restrains the allowed representations. In Section 3,
we use as example lepton and electroweak fields, expressed in the
(5+1)-d space; in Section 4, their the gauge-invariant ${\rm SU(2)_{L}\times U(1)_{Y}}$
interactive theory is formulated, and its Lagrangian compared with
the standard one. We concentrate on the W-lepton vertex contribution;
we find the correct phases and coefficients in a projection operator
that allow for this equality. In Section 5, we summarize relevant
points in the paper.

\section{Gamma-matrix symmetry classification}

In this section, we summarize the main points in the classification
of states and symmetries. More details may be found in Ref. \cite{Besprosvany:2014lwa}.
A massless Dirac equation formulated over the matrix $\Psi$ (and
corresponding conjugate equation)
\begin{eqnarray}
i\gamma_{0}\partial_{\mu}\gamma^{\mu}\Psi={0},\label{JaimeqDi}
\end{eqnarray}
may be used as framework for the classification of states and operators
in an extended space,\footnote{We assume throughout $\hbar=c=1$, and 4-d diagonal metric elements
$g_{\mu\nu}=(1,-1,-1,-1).$} and study symmetry transformations. It also generates free-particle
fermion and bosons on the extended space. Appropriate transformation
operators $U$ acting on field states $\Psi$ can generically be characterized
by the expression
\begin{eqnarray}
\Psi\rightarrow U\Psi U^{\dagger}.\label{transfoGen}
\end{eqnarray}
for both Lorentz and scalar symmetries. In the massive case, some
symmetries are broken, leading to effects as fermion-mass hierarchy
generation, treated elsewhere\cite{Besprosvany:2014lwa}.

The dot product between the elements $\Psi_{a}$, $\Psi\textbf{}_{b}$
can be defined using the trace
\begin{eqnarray}
{\rm tr\ }\Psi_{a}^{\dagger}\Psi_{b}.\label{dotprod}
\end{eqnarray}

An operator $Op$ within this space characterizes a state $\Psi$
with the eigenvalue rule
\begin{eqnarray}
[Op,\Psi]=\lambda\Psi,\label{OpAct}
\end{eqnarray}
consistent with the hole interpretation, and anticipating a second-quantization
description. For example, a boson may be constructed by two fermion
components with positive frequencies $\psi_{1}(x)$, $\bar{\psi}_{2}(x)$
through $\psi_{1}(x)\bar{\psi}_{2}(x)$, with $\bar{\psi}_{2}(x)$
describing an antiparticle.

Eq. \ref{JaimeqDi}, keeping $\mu=0,...,3,$ is assumed within the
larger Clifford algebra, here also understood as a matrix space: $\{\gamma_{\eta},\gamma_{\sigma}\}=2g_{\eta\sigma}$,
$\eta,\sigma=0,...3,5,...,N$, with $N$ the (assumed even) dimension,
whose structure is helpful in classifying the available symmetries
$U$, and solutions $\Psi$, both represented by $2^{N/2}\times2^{N/2}$
matrices. The 4-d Lorentz symmetry is maintained, and uses the generators
\begin{eqnarray}
\sigma_{\mu\nu}=\frac{i}{2}[\gamma_{\mu},\gamma_{\nu}],\label{sigmamunu}
\end{eqnarray}
where $\mu,\nu=0,...,3.$ $U$ contains also $\gamma_{a}$, $a=5,...,N$,
and their products as possible symmetry generators. % CHECK POSITIONIndeed,
the latter elements are scalars for they commute with the Poincar\'e
generators, which contain $\sigma_{\mu\nu}$, and they are also symmetry
operators of the massless Eq. \ref{JaimeqDi}, bilinear in $\gamma_{\mu},$
$\mu=0,...,3$ which is not necessarily the case for mass terms (containing
$\gamma_{0}$). In addition, their products with
\begin{eqnarray}
\tilde{\gamma}_{5}=-i\gamma_{0}\gamma_{1}\gamma_{2}\gamma_{3}\label{tildegamma5}
\end{eqnarray}
are Lorentz pseudoscalars, as $[\tilde{\gamma}_{5},\gamma_{a}]=0$.

The operator algebra was described in Refs. \cite{Besprosvany:2002tv}
and \cite{Besprosvany:2010zz}. In accordance with the above symmetry
generators that emerge from the Clifford algebra ${\mathcal{C}}_{N}$,
for given dimension $N$, any matrix element representing a state
is obtained by combinations of products of one or two $\gamma_{\mu}$,
and elements of the algebra generated by $\gamma_{a}$, $a=5,...,N$,
which define, respectively, their Lorentz (as for 4-d) and scalar-group
representation ${\mathcal{S}}_{N-4}$. A 4-d Clifford matrix subalgebra
is obtained, implying spinor up to bi-spinor elements, thus vectors
and scalar fields, can be described. There is a finite number of partitions
on the matrix space for the states and symmetry operators, consistent
with Lorentz symmetry. These variations are defined by a projection
operators ${\mathcal{P}}_{P}$ with $[{\mathcal{P}}_{P},{\mathcal{P}}_{S}]=0$;
${\mathcal{P}}_{P}$ acts on the Lorentz generator
\begin{eqnarray}
{\mathcal{P}}_{P}[\frac{1}{2}\sigma_{\mu\nu}+i(x_{\mu}\partial_{\nu}-x_{\nu}\partial_{\mu})],\label{Lorentgen}
\end{eqnarray}
and ${\mathcal{P}}_{S}$ on the symmetry operator space leading to
projected scalar generators $I_{a}={\mathcal{P}}_{S}I_{a}$, so that
they determine, respectively, the Poincar\'e generators and the scalar
groups.

The application of these operators follows the operator rule in Eq.
\ref{OpAct}, which assigns states to particular Lorentz and scalar
group representations. For simplicity, we assume ${\mathcal{P}}_{P}={\mathcal{P}}_{S}\neq1$,
as other possibilities are less plausible\cite{Besprosvany:2002tv}.
Thus, the Lorentz or scalar operators act trivially on one side of
solutions of the form $\Psi={\mathcal{P}}_{P}\Psi(1-{\mathcal{P}}_{P})$,
since $(1-{\mathcal{P}}_{P}){\mathcal{P}}_{P}=0$, leading to spin-1/2
states or states belonging to the fundamental representation of the
non-Abelian symmetry groups, respectively.

\begin{figure}
\subfloat[]{\noindent \begin{centering}
\begin{tikzpicture}[scale=1,every node/.style={scale=0.7}]

\draw[thick] (0,0) rectangle (6,6);

\draw[thick] (3,0) -- (3,6);
\draw[thick] (0,3) -- (6,3);

\draw (0.75,0) -- (0.75,6);
\draw (0,5.25) -- (6,5.25);

\node at (0.375,5.625) {(*)};
\node at (1.75,4.25) {${\mathcal S}_{(N-4)R}\times {\mathcal C}_{4}$};
\node at (4.5,1.5) {${\mathcal S}_{(N-4)L}\times {\mathcal C}_{4}$};

\end{tikzpicture}
\par\end{centering}

}\hfill{}\subfloat[]{\noindent \begin{centering}
\begin{tikzpicture}[scale=1,every node/.style={scale=0.7}]

\draw[thick] (0,0) rectangle (6,6);

\draw[thick] (3,0) -- (3,6);
\draw[thick] (0,3) -- (6,3);

\draw (0.75,0) -- (0.75,6);
\draw (0,5.25) -- (6,5.25);

\node at (0.375,5.625) {(*)};
\node at (1.75,5.625) {${F}$};
\node at (4.5,5.625) {${F}$};
\node at (1.75,4.25) {${V}$};
\node at (4.5,1.5) {${V}$};
\node at (0.375,4.25) {${F}$};
\node at (0.375,1.5) {${F}$};
\node at (4.5,4.25) {${S,A}$};
\node at (1.75,1.5) {${S,A}$};

\end{tikzpicture}
\par\end{centering}

}

\begin{centering}
\caption{(a) shows the arrangement of symmetry operators $U$ in matrix space
of arbitrary dimension $N$, after projection over ${\mathcal{S}}_{P}$,
with left-handed and right-handed operators subspaces\cite{Besprosvany:2002tv};
({*}) represents the matrix subspace containing the projector $1-{\mathcal{P}}_{S}=1-{\mathcal{P}}_{P}$;
its choice within the right-handed symmetry components is arbitrary.
(b) shows the arrangement of matrix solutions $\Psi$ in the extended-spin
model is divided into four $\frac{N}{2}\times\frac{N}{2}$ matrix
blocks, containing fermion (F), vector (and axial-) (V), and scalar
(and pseudo-), and antisymmetric (S,A) terms. }

\par\end{centering}

\end{figure}

\noindent In Figure 1(a), presented also in Ref. \cite{Besprosvany:2014lwa},
we show schematically the organization of the symmetry operators,
producing corresponding Lorentz and scalar generators. Fig. 1{(b)}
also depicts the resulting solution representations, distributed according
to their Lorentz classification: fermion, scalar, vector, and antisymmetric
tensor. The matrices are classified according to the chiral projection
operators $\frac{1}{2}(1\pm\tilde{\gamma}_{5})$, leading to $N/2\times N/2$
matrix blocks in ${\mathcal{C}}_{N}$. The space projected by ${\mathcal{P}}_{P}={\mathcal{P}}_{S}\neq1$
is also depicted. Specific combinations also emerge, corresponding
to spin-1/2-fundamental and vector-adjoint, Lorentz and scalar groups
representations, respectively; graphically, scalar-group elements
and vectors occupy the same matrix spots.

In the next Section, we generalize these fields.

\section{(5+1)-dimensional representations}

We review the (5+1)-dimensional representations, which reproduce a
standard-model lepton electroweak sector\cite{Besprosvany:2002py};
one of its coupling terms will be analyzed in the next Section.

\subsection{Fields' construction}

As derived in Section 2, it is possible to write fundamental fields
using as basis matrix products conformed of Lorentz and scalar group
representations. Indeed, the commuting property of the respective
degrees of freedom allows for states and operators to be written as
a product of matrices belonging to the 4-d ${\mathcal{C}}_{4}$, and
matrices within ${\mathcal{S}}_{N-4}$ projected by ${\mathcal{P}}_{S}$;
explicitly, $\Psi=M_{1}M_{2}$, where
\begin{eqnarray}
M_{1}\in{\mathcal{C}}_{4}\ \ {\rm and}\ \ M_{2}\in{\mathcal{P}}_{S}{\mathcal{S}}_{N-4}.\label{Separation}
\end{eqnarray}
An expression with elements of each set is possible through their
passage to each side, using commutation or anticommutation rules.

In the presence of interactions, free fields as generated by Eq. \ref{JaimeqDi},
give way to more general expressions of interactive fermion and boson
fields, keeping their transformation properties:
\begin{description}
\item [{ Vector field}]
\begin{eqnarray}
A_{\mu}^{a}(x)\gamma_{0}\gamma^{\mu}I_{a},\label{AmuExpandfinal}
\end{eqnarray}

\end{description}
where $\gamma_{0}\gamma_{\mu}\in{\mathcal{C}}_{4}$, $I_{a}\in{\mathcal{P}}_{S}{\mathcal{S}}_{N-4}$
is a generator of a given unitary group, according to the projection
operator ${\mathcal{P}}_{S}$.
\begin{description}
\item [{  Fermion field}]
\begin{eqnarray}
\psi_{\alpha}^{a}(x)L^{\alpha}P_{F}M_{a}^{F},\label{Fermion}
\end{eqnarray}

\end{description}
where $M_{a}^{S}$,$M_{a}^{F}\in{\mathcal{P}}_{S}{\mathcal{S}}_{N-4}$
are, respectively, scalar and fermion components, and $L^{\alpha}$
represents a spin component; for example, $L^{1}=(\gamma_{1}+i\gamma_{2})$,
$P_{F}$ is a projection operator of the type in Eq. \ref{Lorentgen},
such that
\begin{eqnarray}
P_{F}\gamma_{\mu}=\gamma_{\mu}P_{F}^{c},\label{transfoyy}
\end{eqnarray}
and we use the complement $P_{F}^{c}=1-P_{F}$, so that a Lorentz
transformation with $P_{F}\sigma_{\mu\nu}$, will describe fermions,
as argued in Section 3; the simplest example for an operator satisfying
such conditions is $P_{F}=(1-\tilde{\gamma}_{5})/2$ \cite{Besprosvany:2002zr,Besprosvany2001},
used by the fermion doublet on Table 1 (see below.) By the argument
after Eq. \ref{Lorentgen}, the fundamental-representation state is
derived from the trivial right-hand action of the operator within
the transformation rule in Eq. \ref{transfoyy}. This means the matrix
entitles spurious ket states contained in the Lorentz-scalar term
$M_{2}$ in Eq. \ref{Separation}.

For the (5+1)-dimensional space, among few choices, ${\mathcal{P}}_{P}=L$,
with $L=\frac{3}{4}-\frac{i}{4}(1+\tilde{\gamma}_{5})\gamma^{5}\gamma^{6}-\frac{1}{4}\tilde{\gamma}_{5}$
is associated to the lepton number, and the resulting symmetry generators
and particle spectrum fits the standard-model electroweak sector.
Specifically, the projected symmetry space also includes the SU(2)$_{L}\times{\rm U(1)_{Y}}$
groups, with respective generators $I_{i}$ and hypercharge $Y$
\begin{eqnarray}
I_{1} & = & \frac{i}{4}(1-\tilde{\gamma}_{5})\gamma^{5}\nonumber \\
I_{2} & = & -\frac{i}{4}(1-\tilde{\gamma}_{5})\gamma^{6}\nonumber \\
I_{3} & = & -\frac{i}{4}(1-\tilde{\gamma}_{5})\gamma^{5}\gamma^{6}\nonumber \\
Y & = & -1+\frac{i}{2}(1+\tilde{\gamma}_{5})\gamma^{5}\gamma^{6}.
\end{eqnarray}
We note that the ${\rm SU(2)}$ generators correctly contain the projection
operator $\frac{1}{2}(1-\tilde{\gamma}_{5})$, confirming the interaction's
chiral nature, which also leads to chiral representations, a feature
that results from nature of the matrix space under projector $L$
and the Lorentz group.

A state basis is presented on Table 1, that contains lepton, as well
as scalar and electroweak vector components; W and Z components are
shown, where the latter normalizations require relative coupling-constant
$g$ and $\frac{g'}{2}$ factors, respectively. Within Eq. \ref{OpAct},
the action of these operators on choices of states $\Psi$ produce
their quantum numbers, also represented. For fermions and vectors,
the second spin component may be obtained from the first by flipping
the spin; e. g., $\nu_{L}^{2}=[L(\gamma_{2}\gamma_{3}-i\gamma_{3}\gamma_{1}),\nu_{L}^{1}]$.

\begin{table}
\noindent \begin{adjustbox}{max width=\textwidth} %
\begin{tabular}{|>{\centering}p{0.15\textwidth}|c|c|c|c|c|c|c|}
\hline
Electroweak multiplet & States $\Psi$ & $\begin{array}{c}
\\
I_{3}\\
\\
\end{array}$ & $\begin{array}{c}
\\
Y\\
\\
\end{array}$ & $\begin{array}{c}
\\
Q\\
\\
\end{array}$ & $\begin{array}{c}
\\
L\\
\\
\end{array}$ & $\begin{array}{c}
\\
\frac{i}{2}L\gamma^{1}\gamma^{2}\\
\\
\end{array}$ & $\begin{array}{c}
\\
L\tilde{\gamma}_{5}\\
\\
\end{array}$\tabularnewline
\hline
\hline
Fermion doublet & $\begin{aligned}\\
\nu_{L}^{1}=\frac{1}{8}(1-\tilde{\gamma}_{5})(\gamma^{0}+\gamma^{3})(\gamma^{5}-i\gamma^{6})\\
\nu_{L}^{2}=\frac{1}{8}(1-\tilde{\gamma}_{5})(\gamma^{0}-\gamma^{3})(\gamma^{5}-i\gamma^{6})\\
e_{L}^{1}=\frac{1}{8}(1-\tilde{\gamma}_{5})(\gamma^{0}+\gamma^{3})(1+i\gamma^{5}\gamma^{6})\\
e_{L}^{2}=\frac{1}{8}(1-\tilde{\gamma}_{5})(\gamma^{0}-\gamma^{3})(1+i\gamma^{5}\gamma^{6})\\
\\
\end{aligned}
$ & $\begin{array}{r}
\\
1/2\\
\\
1/2\\
\\
-1/2\\
\\
-1/2\\
\\
\end{array}$ & $\begin{array}{c}
\\
-1\\
\\
-1\\
\\
-1\\
\\
-1\\
\\
\end{array}$ & $\begin{array}{c}
\\
0\\
\\
0\\
\\
-1\\
\\
-1\\
\\
\end{array}$ & $\begin{array}{c}
\\
1\\
\\
1\\
\\
1\\
\\
1\\
\\
\end{array}$ & $\begin{array}{r}
\\
1/2\\
\\
-1/2\\
\\
1/2\\
\\
-1/2\\
\\
\end{array}$ & $\begin{array}{c}
\\
-1\\
\\
-1\\
\\
-1\\
\\
-1\\
\\
\end{array}$\tabularnewline
\hline
Fermion singlet & $\begin{aligned}\\
e_{R}^{1}=\frac{1}{8}(1+\tilde{\gamma}_{5})\gamma^{0}(\gamma^{0}+\gamma^{3})(\gamma^{5}-i\gamma^{6})\\
e_{R}^{2}=\frac{1}{8}(1+\tilde{\gamma}_{5})\gamma^{0}(\gamma^{0}-\gamma^{3})(\gamma^{5}-i\gamma^{6})\\
\\
\end{aligned}
$ & $\begin{array}{c}
\\
0\\
\\
0\\
\\
\end{array}$ & $\begin{array}{c}
\\
-2\\
\\
-2\\
\\
\end{array}$ & $\begin{array}{c}
\\
-1\\
\\
-1\\
\\
\end{array}$ & $\begin{array}{c}
\\
1\\
\\
1\\
\\
\end{array}$ & $\begin{array}{r}
\\
1/2\\
\\
-1/2\\
\\
\end{array}$ & $\begin{array}{c}
\\
1\\
\\
1\\
\\
\end{array}$\tabularnewline
\hline
Scalar doublet & $\begin{aligned}\\
\frac{1}{4\sqrt{2}}(1-\tilde{\gamma}_{5})\gamma^{0}(1-i\gamma^{5}\gamma^{6})\\
\frac{1}{4\sqrt{2}}(1-\tilde{\gamma}_{5})\gamma^{0}(\gamma^{5}+i\gamma^{6})\\
\\
\end{aligned}
$ & $\begin{array}{r}
\\
1/2\\
\\
-1/2\\
\\
\end{array}$ & $\begin{array}{c}
\\
1\\
\\
1\\
\\
\end{array}$ & $\begin{array}{c}
\\
1\\
\\
0\\
\\
\end{array}$ & $\begin{array}{c}
\\
0\\
\\
0\\
\\
\end{array}$ & $\begin{array}{c}
\\
0\\
\\
0\\
\\
\end{array}$ & $\begin{array}{c}
\\
-2\\
\\
-2\\
\\
\end{array}$\tabularnewline
\hline
Vector singlet & $\begin{gathered}\\
\frac{1}{2\sqrt{2}}\gamma^{0}(\gamma^{1}+i\gamma^{2})Y\\
\frac{1}{2}\gamma^{0}\gamma^{3}Y\\
\frac{1}{2\sqrt{2}}\gamma^{0}(\gamma^{1}-i\gamma^{2})Y\\
\\
\end{gathered}
$ & $\begin{array}{c}
0\\
\\
0\\
\\
0
\end{array}$ & $\begin{array}{c}
0\\
\\
0\\
\\
0
\end{array}$ & $\begin{array}{c}
0\\
\\
0\\
\\
0
\end{array}$ & $\begin{array}{c}
0\\
\\
0\\
\\
0
\end{array}$ & $\begin{array}{r}
1\\
\\
0\\
\\
-1
\end{array}$ & $\begin{array}{c}
0\\
\\
0\\
\\
0
\end{array}$\tabularnewline
\hline
Vector triplet & $\begin{gathered}\\
\frac{1}{8}(1-\tilde{\gamma}_{5})\gamma^{0}(\gamma^{1}+i\gamma^{2})(\gamma^{5}-i\gamma^{6})\\
\frac{1}{4\sqrt{2}}(1-\tilde{\gamma}_{5})\gamma^{0}(\gamma^{1}+i\gamma^{2})\gamma^{5}\gamma^{6}\\
\frac{1}{8}(1-\tilde{\gamma}_{5})\gamma^{0}(\gamma^{1}+i\gamma^{2})(\gamma^{5}+i\gamma^{6})\\
\\
\end{gathered}
$ & $\begin{array}{r}
1\\
\\
0\\
\\
-1
\end{array}$ & $\begin{array}{c}
0\\
\\
0\\
\\
0
\end{array}$ & $\begin{array}{r}
1\\
\\
0\\
\\
-1
\end{array}$ & $\begin{array}{c}
0\\
\\
0\\
\\
0
\end{array}$ & $\begin{array}{c}
1\\
\\
1\\
\\
1
\end{array}$ & $\begin{array}{c}
0\\
\\
0\\
\\
0
\end{array}$\tabularnewline
\hline
\end{tabular}

\end{adjustbox}

\noindent \caption{Massless fermion and boson states in (5+1)-d extension, momentum along
${\pm{\bf {\hat{z}}}}$, with projection given by the lepton number
${\mathcal{P}}_{P}=L$, under the operators SU(2)$_{L}$ $I_{3}$
component, hypercharge $Y$, charge $Q=I_{3}+\frac{1}{2}Y$, lepton
operator $L$, spin projection $\frac{i}{2}L\gamma^{1}\gamma^{2}$,
and chirality $L\tilde{\gamma}_{5}$ (the coordinate dependence is
omitted.) }
\end{table}

Ref. \cite{Besprosvany:2002py} set thumb rules to derive some gauge-invariant
terms, identifying elements between the extended-spin space and standard
Lagrangian terms. Ref. \cite{Besprosvany:2014lwa} formally translated
the field information that emerges from the extended-spin space, to
derive an interactive gauge theory. Next, we show for the lepton-W
vertex the workings of the equivalence between the extended-spin model
and the standard Lagrangian formulation.

\section{Fermion-W electroweak Lagrangian}

The fields within the extended-spin basis can be used to construct
a standardly-formulated Lagrangian. This amounts to using elements
with a well-defined group structure to get Lorentz-scalar gauge-invariant
combinations. Choosing scalar elements that result from the direct
product in Eq. \ref{dotprod}, one obtains an interactive theory,
as the same particle content is maintained.

Indeed, a gauge-invariant fermion-vector interaction term results,
constructing matrix elements containing the vector field, together
with fermion, with input from Eqs. \ref{AmuExpandfinal}-\ref{Fermion},
by taking the trace. Invariant elements are obtained adding to the
fermion free Lagrangian (that implies the Dirac equation \ref{JaimeqDi})
the vector contribution in Eq. \ref{AmuExpandfinal}. The latter extracts
the identity-matrix coefficient, leading to the usual Lagrangian components.
A general fermion-vector component is
\begin{eqnarray}
\frac{1}{N_{f}}{\rm tr}\Psi^{\dagger}\{[i\partial_{\mu}I_{den}+gA_{\mu}^{a}(x)I_{a}]\gamma_{0}\gamma^{\mu}-M\gamma_{0}\}\Psi P_{f},\label{Bilin}
\end{eqnarray}
where $\Psi$ is a field representing in this case spin-1/2 particles.
$I_{a}$ is the group generator in a given representation, $g$ is
the coupling constant, $N_{f}$ contains the normalization (and similar
terms below), and $I_{den}$ the identity scalar group operator in
the same representation (which will be omitted hence). $M$ is generally
a mass operator whose restrictions provide information on fermion
masses\cite{Besprosvany:2002py}, \cite{Besprosvany:2014lwa}. The
operator $P_{f}$ is introduced to avoid cancelation of non-diagonal
fermion elements. Such an operator is necessary because of spurious
left ket components of fermions in $\Psi$. For example,
\begin{eqnarray}
P_{f}=\frac{1}{\sqrt{2}}(\tilde{\gamma}^{5}-\gamma^{0}\gamma^{1})\label{Pf}
\end{eqnarray}
as $[P_{f},L]=[P_{f},(1-\tilde{\gamma}_{5})L]=0$, provides a non-trivial
combination with the correct quantum numbers for the fermion pair
$\Psi_{a}P_{f}\Psi_{b}^{\dagger}$ (with $\Psi_{a},\Psi_{b}$ either
doublet or singlet fermions, on Table 1), and maintains their normalization,
spin, lepton and electroweak representation.

The invariance under transformations in Eq. \ref{transfoGen} can
be verified independently, using the separation in Eq. \ref{Separation}
into Lorentz and scalar symmetries; under Lorentz and gauge-group
transformations of the extended-spin space\cite{Besprosvany:2014lwa}.
Eq. \ref{Bilin} is invariant under the Lorentz transformation, provided
the vector field transforms as
\begin{eqnarray}
A_{\mu}^{a}(x)I_{a}\rightarrow\Delta_{\mu}^{\ \ \nu}A_{\nu}^{a}(x)I_{a},\label{transformsLor}
\end{eqnarray}
where we use the identity relating the spin representation of the
Lorentz group in
\begin{eqnarray}
U\gamma^{\mu}U^{-1}=({\Delta^{-1}})_{\ \ \nu}^{\mu}\gamma^{\nu},\label{LorentzIden}
\end{eqnarray}
and $\Delta_{\ \ \nu}^{\mu}$ is a $4\times4$ Lorentz matrix transforming
coordinates as $x^{\mu}\rightarrow\Delta_{\ \ \nu}^{\mu}x^{\nu}$.
The equation is also invariant under the local transformation, under
the condition the vector field transforms as
\begin{eqnarray}
A_{\mu}^{a}(x)I_{a}\rightarrow UA_{\mu}^{a}(x)I_{a}U^{\dagger}-\frac{i}{g}(\partial_{\mu}U)U^{\dagger},\label{transforms}
\end{eqnarray}

Thus, the fermion-vector Lagrangian in Eq. \ref{Bilin} with the fields
on Table 1, leads to the fermion electroweak standard-model Lagrangian
contribution\cite{Glashow:1961tr,Weinberg:1967tq}, also derived heuristically
in Refs. \cite{Besprosvany:2002py} and \cite{Besprosvany:2010zz},
\begin{eqnarray}
{\bar{\boldmathPsi_{l}}}[i\partial_{\mu}+\frac{1}{2}g\tau^{a}W_{\mu}^{a}(x)-\frac{1}{2}g'B_{\mu}(x)]\gamma^{\mu}{\boldmathPsi_{l}}+\nonumber \\
{\bar{\psi_{r}}}[i\partial_{\mu}-g'B_{\mu}(x)]\gamma^{\mu}{\psi_{r}},
\end{eqnarray}
which contains a left-handed hypercharge $Y_{l}=-1$ SU(2) doublet
\begin{eqnarray}
\boldmathPsi_{l}(x)=\left(\begin{array}{lcr}
\nu_{L}(x)\\
e_{L}(x)
\end{array}\right),\label{doubletconve}
\end{eqnarray}
with two polarization components as, e. g.,
\begin{eqnarray}
\nu_{L}(x)=\left(\begin{array}{lcr}
\psi_{\nu L}^{1}(x)e^{ip_{\nu L1}(x)}\\
\psi_{\nu L}^{2}(x)e^{ip_{\nu L2}(x)}
\end{array}\right),\label{spinorconve}
\end{eqnarray}
and we choose polar coordinates; a right-handed $Y_{r}=-2$ singlet
$\psi_{r}$, with likewise notation, and the corresponding gauge-group
vector bosons and coupling constants, $B_{\mu}(x)$, $W_{\mu}^{a}(x)$,
and $g$, $g'$, respectively.

In the following, we justify $P_{f}$ in Eq. \ref{Pf}, showing the
equivalence of the spin-extended W-fermion vertex containing the operator
$W_{\mu}^{i}(x)\gamma_{0}\gamma^{\mu}$ (the $\frac{1}{2}g$ factor
hence omitted), within Eq. \ref{Bilin}, in comparison to the conventional
expression in Eq. \ref{conventional}.

The $(5+1)-$d space allows for charge $0$ and $-1$ components,
associated to lepton (neutrino and electron) fields
\begin{eqnarray}
\boldmathPsi_{L}^{l}(x)=\sum_{\alpha}\left(\begin{array}{lcr}
\psi_{\nu L}^{\alpha}(x)e^{ip_{\nu L\alpha}(x)}\nu_{L}^{\alpha}\\
\psi_{eL}^{\alpha}(x)e^{ip_{eL\alpha}(x)}e_{L}^{\alpha}
\end{array}\right),
\end{eqnarray}
\begin{eqnarray}
{\Psi_{R}^{e}}(x)=\sum_{\alpha}\psi_{eR}^{\alpha}(x)e^{ip_{eR\alpha}(x)}e_{R}^{\alpha},
\end{eqnarray}
and the spinor-lepton components shown in Table 1 (see notation).
The conventional states are assumed real and obtained within the $\gamma^{\mu}$-Dirac
representation. We also choose polar coordinates for the components
to pinpoint phase effects.

The relevant (5+1)-d projection terms are
\begin{eqnarray}
\!\!\!\!\!\!\!\!\!\!\!\!\!\!\!\!P_{f}=\tilde{g}_{5}{\tilde{\gamma}_{5}}+g_{I}I+g_{01}\gamma^{0}\gamma^{1}+g_{02}\gamma^{0}\gamma^{2}+\!\!\!\!\!\!\nonumber \\
\!\!\!\!\!\!\!\!\!\!\!\!\!\!\!\!g_{03}\gamma^{0}\gamma^{3}+g_{12}\gamma^{1}\gamma^{2}+g_{13}\gamma^{1}\gamma^{3}+g_{23}\gamma^{2}\gamma^{3}+\!\!\!\!\!\!\nonumber \\
\!\!\!\!\!\!\!\!\!\!\!(g_{\tilde{5}56}\tilde{\gamma}_{5}+g_{I56}I+g_{0156}\gamma^{0}\gamma^{1}+g_{0256}\gamma^{0}\gamma^{2}+g_{0356}\gamma^{0}\gamma^{3}+\!\!\!\!\!\!\nonumber \\
\!\!\!\!\!\!\!\!\!\!\!\!\!\!\!\!\!g_{1256}\gamma^{1}\gamma^{2}+g_{1356}\gamma^{1}\gamma^{3}+g_{2356}\gamma^{2}\gamma^{3})\gamma^{5}\gamma^{6}.\ \
\end{eqnarray}

%Phases: $e^{i p_{eL1}}$ $e^{i p_{eL2}}$ $e^{i p_{\nu L1   }}$ $e^{i p_{\nu L2}}$ $e^{i p_{eR1}}$ $e^{i %p_{eR2}}$

For the extended-spin model with $\boldmathPsi_{L}^{l}(x)$, the coefficient
of the $e_{L}^{1}$ associated term $(\psi_{L}^{1}(x))^{2}$ is

$(A-B)[W_{0}^{3}(x)-W_{3}^{3}(x)]$

$A=\frac{1}{2}(g_{I}+g_{\tilde{5}}-ig_{\tilde{5}56}-ig_{56})$

$B=-\frac{1}{2}(g_{03}-ig_{12}-ig_{0356}-g_{1256}).$

For the conventional term with $\boldmathPsi_{l}(x)$, the $e_{L}^{1}$
coefficient is

$\frac{1}{2}[W_{0}^{3}(x)-W_{3}^{3}(x)]$.

For the extended-spin model with $\boldmathPsi_{L}^{l}(x)$, the coefficient
of the ${e_{L}}^{2}$ associated term $(\psi_{eL}^{2}(x))^{2}$ is

$(A+B)[W_{0}^{3}(x)+W_{3}^{3}(x)].$

For the conventional term with $\boldmathPsi_{l}(x)$, the $e_{L}^{2}$
coefficient is

$\frac{1}{2}[W_{0}^{3}(x)+W_{3}^{3}(x)].$

We conclude the choice $A=1/2$, $B=0$ equates the two expressions.
Given the expressions for $A$, $B$, there is some freedom in the
coefficients $g_{i}$ choice. These terms do not provide phase information,
unlike cross terms:

Indeed, for the extended-spin model with $\boldmathPsi_{L}^{l}(x)$,
the $\nu_{L}^{1}$ $e_{L}^{2}$ coefficients of the associated term
$\psi_{\nu L}^{1}(x)\psi_{eL}^{2}(x)$ are presented for each $W_{\mu}^{i}(x)$:

\[
\begin{array}{l}
\!\!\!\!\!\!W_{1}^{1}(x)\\
-\frac{1}{2}ie^{-i({p_{eL2}}(x)+{p_{\nu L1}}(x))}\left[{C}\left(e^{2i{p_{eL2}}(x)}+e^{2i{p_{\nu L1}}(x)}\right)+\right.\\
\left.{D}\left(e^{2i{p_{\nu L1}}(x)}-e^{2i{p_{eL2}}(x)}\right)\right]\\
\!\!\!\!\!\!W_{2}^{2}(x)\\
-\frac{1}{2}ie^{-i({p_{eL2}}(x)+{p_{\nu L1}}(x))}\left[{C}\left(e^{2i{p_{eL2}}(x)}+e^{2i{p_{\nu L1}}(x)}\right)+\right.\\
\left.{D}\left(e^{2i{p_{\nu L1}}(x)}-e^{2i{p_{eL2}}(x)}\right)\right]\\
\!\!\!\!\!\!W_{1}^{2}(x)\\
\frac{1}{2}e^{-i({p_{eL2}}(x)+{p_{\nu L1}}(x))}\left[{C}\left(e^{2i{p_{eL2}}(x)}-e^{2i{p_{\nu L1}}(x)}\right)-\right.\\
\left.{D}\left(e^{2i{p_{\nu L1}}(x)}+e^{2i{p_{eL2}}(x)}\right)\right]\\
\!\!\!\!\!\!W_{2}^{1}(x)\\
\frac{1}{2}e^{-i({p_{eL2}}(x)+{p_{\nu L1}}(x))}\left[{-C}\left(e^{2i{p_{eL2}}(x)}-e^{2i{p_{\nu L1}}(x)}\right)+\right.\\
\left.{D}\left(e^{2i{p_{\nu L1}}(x)}+e^{2i{p_{eL2}}(x)}\right)\right],
\end{array}
\]
where $C=g_{01}-ig_{23}-ig_{0156}-g_{2356}$

$D=-ig_{02}+g_{13}-g_{0256}-ig_{1356}.$

For the conventional term with $\boldmathPsi_{l}(x)$, the $\nu_{L}^{1}$
$e_{L}^{2}$ coefficient is
\begin{multline*}
\frac{1}{2}e^{-i({p_{eL2}(x)}+{p_{\nu L1}(x)})}\left[i\left(e^{2i{p_{eL2}(x)}}+e^{2i{p_{\nu L1}}}\right)(W_{1}^{1}(x)+W_{2}^{2}(x))+\right.\\
\left.\left(e^{2i{p_{eL2}(x)}}-e^{2i{p_{\nu L1}(x)}}\right)(W_{2}^{1}(x)-W_{1}^{2}(x))\right].
\end{multline*}
We conclude the choice $C=-1$, $D=0$ matches both terms.

While the expression in Eq. \ref{Pf} is consistent with the above
$A$, $B$, $C$, $D$ values (with overall factor linked to the normalization
$N_{f}$ in Eq. \ref{Bilin}), we highlight that a freedom exists
for other $P_{f}$ choices.

Finally, this comparison was carried out under a $\gamma$-matrix
choice that leads to a basis as in Eqs. \ref{doubletconve}, \ref{spinorconve}.
This required fixing the phases, to complete the identification of
states. The phases are given as (to be put on Table 1 states): $e_{L}^{1}\rightarrow-ie_{L}^{1}$
and $e_{L}^{2}\rightarrow-ie_{L}^{2}$. One can check that this solution
fits all other terms.

\section{Conclusions}

This paper dealt with translating a previously proposed standard-model
extension, the spin-extended model, to a Lagrangian formalism, showing
the correspondence of its generated Lagrangian with that of the standard
model, making a specific comparison with one of its components. The
final objective is to use the model's restrictions to obtain standard-model
information.

We first made a brief introduction to the model, highlighting its
main features, and quoting relevant information it generated in previous
references. A matrix space is used in which both symmetry generators
and fields are formulated. For given dimension, a chosen non-trivial
projection operator ${\mathcal{P}}_{P}$ constrains the matrix space,
determining the symmetry groups, and the arrangement of fermion and
boson representations. In particular, spin-1/2, and 0 states are obtained
in the fundamental representation of scalar groups and spin-1 states
in the adjoint representation. After expressing fields within this
basis, a gauge-invariant field theory is constructed, based on the
Lorentz and obtained scalar symmetries.

%first reviews the construction of  an extended spin  space.
In comparison with Ref. \cite{Besprosvany:2014lwa}, in which formal steps were
carried out that relate the spin-extended model with the standard
model, here we examine in detail two associated Lagrangian expressions,
and extract information on the conditions for which they match. The
term-by-term comparison shows special features: one is a need to fix
phases, and the second is the freedom in the choice of the projection
operator, all of which teaches how to match the two formalisms.

As it turns out, the Lagrangian fitting of the projection operator
and the phases, done for the W-lepton term in (5+1)-d, is enough to
show the equivalence of the rest of the other components, as the kinetic
term, and other vertices.

Given the formalization level achieved by the spin-extended model,
it is relevant to mention other of its advantages, as compared with
other extensions. In particular, the chiral property of the model's
fermion representations contrasts with the difficulty to reproduce
it in traditional extensions as the Kaluza-Klein theory. Moreover,
while a grand-unified group limits the representations among which
particles are chosen, in our case, the representations are determined
by the chosen dimension and projection operator over the space. In
fact, the specific combinations (spin-1/2)-fundamental and vector-adjoint
are derived, matching the Lorentz scalar groups representations, respectively;
graphically, vectors and scalars group elements occupy the same places
in the array of extended space of spin), as shown in Fig 1.

The question about what sets the dimension of this extension to derive
groups and representations of the standard model, equally applies
to strings, as there is an infinite number of possible groups that
contain the standard model. The answer for both extensions depends
on whether low dimension numbers give relevant information, and on
predictability, as in our case, in which derived features such as
chiral SU$_{L}$(2) representations.

Although the extensions of the standard model provide additional information
about it, many mysteries remain unsolved. With its bottom-up approach,
this model reduces the possibilities of groups and representations
to describe the particles and their quantum numbers, in contrast,
e. g., with those available in string theory, with its multiplicity
of representation and compactfication choices.

%The complete SM interactions are faithfully described in the  (9+1)-d case, including also fermions %interaction\cite{bespro9m1}. Other obtained features of an extended spin space, as  the coupling constant are derived %elsewhere\cite{besprocoupli}.

The paper's standard-model extension satisfies basic requirement of
correct symmetries, including Lorentz and gauge ones, description
of standard-model particles, and field-theory formulation, in addition
to its standard-model prediction provision (the latter two is what
the paper deals with.) This supports the view that it is an extension
worth considering.

%as its basic building block, should allow for a generalization and%application within the Kaluza-Klein framework, and in  theories such as supersymmetry$\cite{wess}$ and%those accounting for gravity, such as supergravity$\cite{wess}$

The spin-extended model throws light on some standard model enigmas.
To the extent that this extension can be translated to the conventional
field-theory formulation of the standard model, which we show in this
paper is possible, it becomes more relevant.

\bibliographystyle{elsart-num}

\begin{thebibliography}{10}
\bibitem{CMS} CMS collaboration, {Observation of a new boson at
a mass of 125 GeV with the CMS experiment at the LHC}, Phys. Lett.
B716 (2012) 30--61. % arXiv:1207.7235. Bibcode:2012PhLB..716...30C. doi:10.1016/j.physletb.2012.08.021.

\bibitem{Atlas} ATLAS collaboration, {Observation of a New Particle
in the Search for the Standard Model Higgs Boson with the ATLAS Detector
at the LHC}, Phys. Lett. B716 (2012) 1--29. %arXiv:1207.7214. Bibcode:2012PhLB..716....1A.

\bibitem{Nambu} Y.~Nambu, G.~Jona-Lasinio, {Dynamical Model of
Elementary Particles Based on an Analogy with Superconductivity. I
}, Phys. Rev. 122 (1962) 345--358.

\bibitem{Shima:2000es} K.~Shima, {Supersymmetric structure of space-time
and matter: Superon - graviton model}, Phys. Lett. B501 (2001) 237--244.

\bibitem{Chisholm:1999pg} J.~Chisholm, R.~Farwell, {Gauge transformations
of spinors within a Clifford algebraic structure}, J. Phys. A32 (1999)
2805--2823.

\bibitem{raey} M.~Pavsic, On the unification of interactions by
Clifford algebra, Advances in Applied Clifford Algebras 20~(3-4)
(2010) 781--801. %\newline\urlprefix\url{http://dx.doi.org/10.1007/s00006-010-0222-z}

\bibitem{Besprosvany:2002zr} J.~Besprosvany, {Gauge and space-time
symmetry unification}, Int. J. Theor. Phys. 39 (2000) 2797--2836.

\bibitem{Coleman:1967ad} S.~R. Coleman, J.~Mandula, {All possible
symmetries of the S matrix}, Phys. Rev. 159 (1967) 1251--1256.

\bibitem{Besprosvany2001} J.~Besprosvany, Electroweak model from
generalized Dirac equation with boson and fermion solutions, Nucl.
Phys. (Proc. Suppl.) B 101 (2001) 323--323.

\bibitem{Besprosvany:2002tv} J.~Besprosvany, {Standard model particles
and interactions from field equations on spin 9+1-dimensional space},
Phys. Lett. B578 (2004) 181--186.

\bibitem{Besprosvany:2002py} J.~Besprosvany, {Electroweakly interacting
scalar and gauge bosons, and leptons, from field equations on spin
5+1 dimensional space}, Int. J. Mod. Phys. A20 (2005) 77--93.

\bibitem{Besprosvany:2014lwa} J.~Besprosvany, R.~Romero, {Representation
of quantum field theory in an extended spin space and fermion mass
hierarchy}, Int. J. Mod. Phys. A29 (2014) 1450144.

\bibitem{Besprosvany:2010zz} J.~Besprosvany, R.~Romero, {Extended
spin symmetry and the standard model}, American Institute of Physics
Conf. Proc. 1323 (2010) 16--27.

\bibitem{Glashow:1961tr} S.~Glashow, {Partial Symmetries of Weak
Interactions}, Nucl. Phys. 22 (1961) 579--588.

\bibitem{Weinberg:1967tq} S.~Weinberg, {A Model of Leptons}, Phys.
Rev. Lett. 19 (1967) 1264--1266.\end{thebibliography}

\expandafter\ifx\csname url\endcsname\relax \global\long\def\url#1{\texttt{#1}}
\fi \expandafter\ifx\csname urlprefix\endcsname\relax\global\long\def\urlprefix{URL }
\fi

\end{document}